\documentclass[journal,transmag]{IEEEtran}
\usepackage{cite}

\ifCLASSINFOpdf
  \usepackage[pdftex]{graphicx}
\else
\fi
\usepackage{amsmath}
\usepackage{braket}
\usepackage{amssymb}
\usepackage{color,cancel}
\newcommand{\Argum}[1]{\ensuremath{\! \left( #1 \right)}}

\newcommand{\TextSub}[1]{\ensuremath{_{\mbox{\tiny#1}}}}

\newcommand{\mc}[1]{\ensuremath{\mathcal{ #1 }}}  

\usepackage[normalem]{ulem}

\begin{document}
%
\title{Tunable, Hardware-based Quantum Random Number Generation using Coupled Quantum Dots}


\author{\IEEEauthorblockN{Heath McCabe, \IEEEauthorrefmark{1}
Scott M.\ Koziol, \IEEEauthorrefmark{1} \IEEEmembership{Member,~IEEE}\\
Gregory L.\ Snider, \IEEEauthorrefmark{2}\IEEEmembership{Fellow,~IEEE} and
Enrique P.\ Blair, \IEEEauthorrefmark{1}~\IEEEmembership{Senior Member,~IEEE}}
\IEEEauthorblockA{\IEEEauthorrefmark{1}Electrical and Computer Engineering Department,
Baylor University, Waco, TX 76798 USA}
\IEEEauthorblockA{\IEEEauthorrefmark{2}Electrical Engineering Department,
University of Notre Dame, Notre Dame, IN 46556 USA}
\thanks{Manuscript received Month DD, 2019; revised Month DD, 2019. 
Corresponding author: E. Blair (email: Enrique\_Blair(at)Baylor.edu).}}

%



\IEEEtitleabstractindextext{%
\begin{abstract}
Random numbers are a valuable commodity in gaming and gambling, simulation, conventional and quantum cryptography, and in non-conventional computing schemes such as stochastic computing. We propose to generate a random bit using a position measurement of a single mobile charge on a coupled pair of quantum dots. True randomness of the measurement outcome is\textemdash depending on implementation\textemdash provided by statistical mechanics, or by quantum mechanics via Born's rule. A random bit string may be generated using a sequence of repeated measurements on the same double quantum dot (DQD) system. Any bias toward a ``0'' measurement or a ``1'' measurement may be removed or tuned as desired simply by adjusting the bias between the dots.  Device tunability provides versatility, enabling this quantum random number generator (QRNG) to support applications in which no bias is desired, or where a tunable bias is desired. We discuss a metal-dot implementation as well as a molecular implementation of this QRNG. Basic quantum mechanical principles are used to study power dissipation and timing considerations for the generation of random bit strings. The DQD offers a small form factor and, in a metallic implementation, is usable in the case where cryogenic operations are desirable (as in the case of quantum computing). For room-temperature applications, a molecular DQD may be used.
\end{abstract}

}

\maketitle

\IEEEdisplaynontitleabstractindextext

%
\IEEEpeerreviewmaketitle

\section{Introduction}

Random numbers are a valuable commodity in many applications including gaming and lotteries, simulation techniques \cite{1949MonteCarlo}, cryptographic schemes \cite{BB84original}, and in probing fundamental questions of quantum mechanics \cite{1998BellViolation}.

Several approaches to generating random numbers exist, and tradeoffs between various figures of merit often must be weighed in selecting a solution suitable to the requirements of the particular application. 
In some cases, pseudorandom numbers\textemdash deterministically generated using either hardware or software approaches\textemdash may be adequate. 
One simple, low-cost hardware approach uses the linear feedback shift register (LFSR) \cite{LFSR-Patent}. A less conventional approach to a pseudorandom number generator (PRNG) uses quantum-dot cellular automata (QCA)\cite{QCA_PRNG}. When PRNGs do not provide adequate information entropy, approximate approaches to randomness, such as entropy-gathering may be used. Yet, even entropy-gathering techniques have vulnerabilities \cite{UNIXrandomness_vulnerability}. When true randomness is desired, stochastic processes may be used to build true random number generators (TRNGs). One example leverages the stochastic nature of memristors \cite{TRNG_memristor}. Still other approaches to true random number generation leverage quantum effects. Measurement in quantum systems is known to be truly stochastic, and this is the basis of some commercially-available TRNGs \cite{IDQuantiqeWhitePaper3.0}.

Random distributions generated using quantum hardware are likely to have a bias due to sytem nonidealities and must be tuned algorithmically \cite{ulam_vneumann_monte_carlo}, but this comes at the cost of latency and additional conventional computational resources. 
In some applications, power, latency, weight, device complexity, or die area come at a premium, and it is desirable to eliminate additional software conditioning of random bit strings. In other applications, such as stochastic computing \cite{StochasticCompSeminal1967}, not only is a bias in random bit strings desirable, but it is also important to be able to dynamically tune the bias. In this case, stochastic numbers are encoded on the mean value of random bit strings, and simple logic such as AND or OR gates are adequate to perform fundamental mathematical calculations such as addition or multiplication. The mean value of the bit strings must be tunable in order to represent various stochastic numbers.

Here, we propose a straight-forward scheme for generating random numbers with a tunable mean using a coupled pair of quantum dots and a mobile charge. This is in contrast to other proposals, which require a larger array of devices \cite{TRNG_memristor, QCA_PRNG}, or are specific to QCA and are not tunable \cite{QCA_PRNG}. In this paper, localized charge states of a single mobile electron encode a bit. Adjusting the detuning between the quantum dots changes the ground state and thus the probabilities of measurement for the localized states. A projective measurement of position in this system, then, is equivalent to a quantum coin toss with a tunable mean. Repeated measurements yield a random bit string, which may be used generally in any application of random numbers, or specifically as a stochastic number in stochastic computing. This discussion begins with implementation-agnostic theory (Section \ref{sect:theory}) of generating single bits and bit strings. Then, two implementations for this system are described in Section \ref{sect:implementations}. A discussion of the bit rates possible for such tunable quantum random number generators is provided. The tunable mean makes this device suitable for applications where no bias toward either binary outcome is desired, or applications such as stochastic computing where a tunable bias is desired. 

\section{Theory} \label{sect:theory}

\subsection{A Quantum System With Tunable Statistics}

\begin{figure}[htbp] 
   \centering
   \includegraphics[width=0.65\columnwidth]{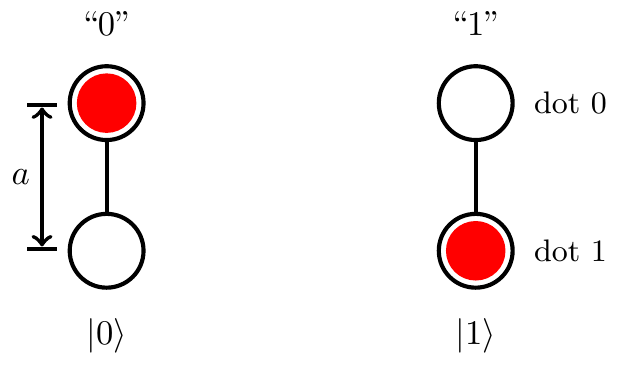} 
   \caption{Localized states of a single electron (red disc) on a coupled pair of quantum dots (black circles) encode two binary states.}
   \label{fig:DoubleDotBasis}
\end{figure}

Consider one mobile electron in a coupled pair of quantum dots, which we refer to as a ``double quantum-dot'' system, or simply a ``DQD.'' The localized electronic states are defined as states $\ket{0}$ and $\ket{1}$, as shown in Fig.\ \ref{fig:DoubleDotBasis}. These charge states provide a basis $\mathcal{B} = \{\ket{0}, \ket{1}\}$ in which the arbitrary quantum superposition state $\ket{\psi}$ may be written as $\ket{\psi} = c_0 \ket{0} + c_1 \ket{1}$ with probability amplitudes $\{ c_0, c_1\}$. While $\ket{\psi}$ may be a quantum superposition of $\ket{0}$ and $\ket{1}$, a position measurement will yield ``0'' or ``1'' only. The probability amplitude $c_x \in \mathbb{C}$ is related to the probability $p \Argum{x}$ of measuring state $\ket{x}$:
\begin{equation}
p \Argum{x} = |c_x|^2 = c_x ^{\ast} c_x \; . \label{eqn:px}
\end{equation}
It is a well-known feature of quantum mechanics that the result of such a measurement is truly random; therefore, a measurement on this DQD system is the quantum generation of a random bit, and we call this system a quantum random number generator (QRNG). Furthermore, for this DQD, the measurement probabilities $p \Argum{k}$ may be dynamically tuned, so we designate this DQD system as a \textit{tunable} QRNG.

We may write the Hamiltonian for the DQD as
\begin{equation}
\hat{H} = -\gamma \hat{\sigma}_x + \frac{\Delta}{2} \hat{\sigma}_z \, ,
\label{eqn:Hamiltonian}
\end{equation}
where $\gamma$ is the tunneling energy between basis states, $\Delta$ is the detuning, $\Delta = \braket{1|\hat{H}|1} - \braket{0|\hat{H}|0}$. Both $\gamma$ and $\Delta$ have units of energy, and the unitless operators  $\hat{\sigma}_{\alpha}$, with $\alpha \in \{ x, y, z\}$, are the Pauli operators:
\begin{align}
\hat{\sigma}_x & = \ket{1} \bra{0} + \ket{0} \bra{1} \, \nonumber  \\
\hat{\sigma}_y & = i \left(  \ket{1} \bra{0} - \ket{0} \bra{1}  \right) \, \mbox{ and}\nonumber \\
\hat{\sigma}_z & = \ket{1} \bra{1} - \ket{0} \bra{0}  \, .
\end{align}

The solutions to the time-independent Schr\"odinger equation
\begin{equation}
\hat{H} \ket{\phi_n} = E_n \ket{\phi_n}
\end{equation}
are the eigenstates $\{ \ket{\phi_n} \}$ of $\hat{H}$ with correpsonding eigenenergies $\{E_n\}$, where $n \in \{ 1, 2\}$ and $E_1 < E_2$.  The eigenenergies may be found symbolically by finding the eigenvalues of the matrix  $H$ which represents $\hat{H}$ in the $\mc{B}$ basis. The eigenvalues may be used to find the eigenvectors of $H$, which represent the eigenstates, $\{\ket{\phi_1}, \ket{\phi_2}\}$. Here, $\ket{\phi_1}$ is the ground state, and the ground state eigenenergy is $E_1$. It can be shown that
\begin{equation}
E_1 =  \braket{\phi_1 | \hat{H} | \phi_1} = -\frac{1}{2} \sqrt{4\gamma^2 +\Delta^2} \; ,
\end{equation}
and 
\begin{equation}
\ket{\phi_1} =  \frac{1}{\sqrt{\alpha^2 + 1}} \left( \alpha \ket{0} + \ket{1} \right) \, ,
\end{equation}
for which
\begin{equation}
\alpha = \frac{\Delta + \sqrt{4\gamma^2 + \Delta^2} }{2\gamma} \, .
\end{equation}
The state $\ket{\phi_2}$ is the excited state with energy $E_2$.

For a DQD relaxed to the ground state $\ket{\phi_1}$, the random position measurement outcomes $x \in \{ 0, 1 \}$ will have the following probabilities, defined over a large ensemble of measurements:
\begin{align}
p\Argum{0} & = \frac{\alpha^2}{\alpha^2 + 1} = \frac{1}{2} \left(1  + \frac{\Delta}{ \sqrt{4 \gamma^2 + \Delta^2}} \right) \label{eqn:p0}\\
p \Argum{1} & = \frac{1}{\alpha^2 + 1} = \frac{1}{2} \left( 1 - \frac{\Delta}{ \sqrt{4 \gamma^2 + \Delta^2}} \right) . \label{eqn:p1}
\end{align}
The mean value, $\bar{x}$, of many measurements then, may be found by performing a weighted sum over all possible outcomes, $x$:
\begin{equation}
\bar{x} = \sum_{x=0}^{1} x p\Argum{x} = p\Argum{1} = \frac{2\gamma^{2}}{\Delta^{2}+\Delta\sqrt{4\gamma^{2}+\Delta^{2}}+4\gamma^{2}} \; .
\end{equation}
Here, $\bar{x}$ describes position measurements over an ensemble of identically-prepared DQDs. Equivalently, $\bar{x}$ could describe a series of measurements on a single DQD system, with the state reset to $\ket{\phi_1}$ prior to each successive measurement. The resetting of the state to $\ket{\phi_1}$ is necessary for this interpretation of $\bar{x}$, and it avoids identical and repeated measurement results via the quantum Zeno effect.

\begin{figure}[htbp] 
   \centering
   \includegraphics[width=1\columnwidth]{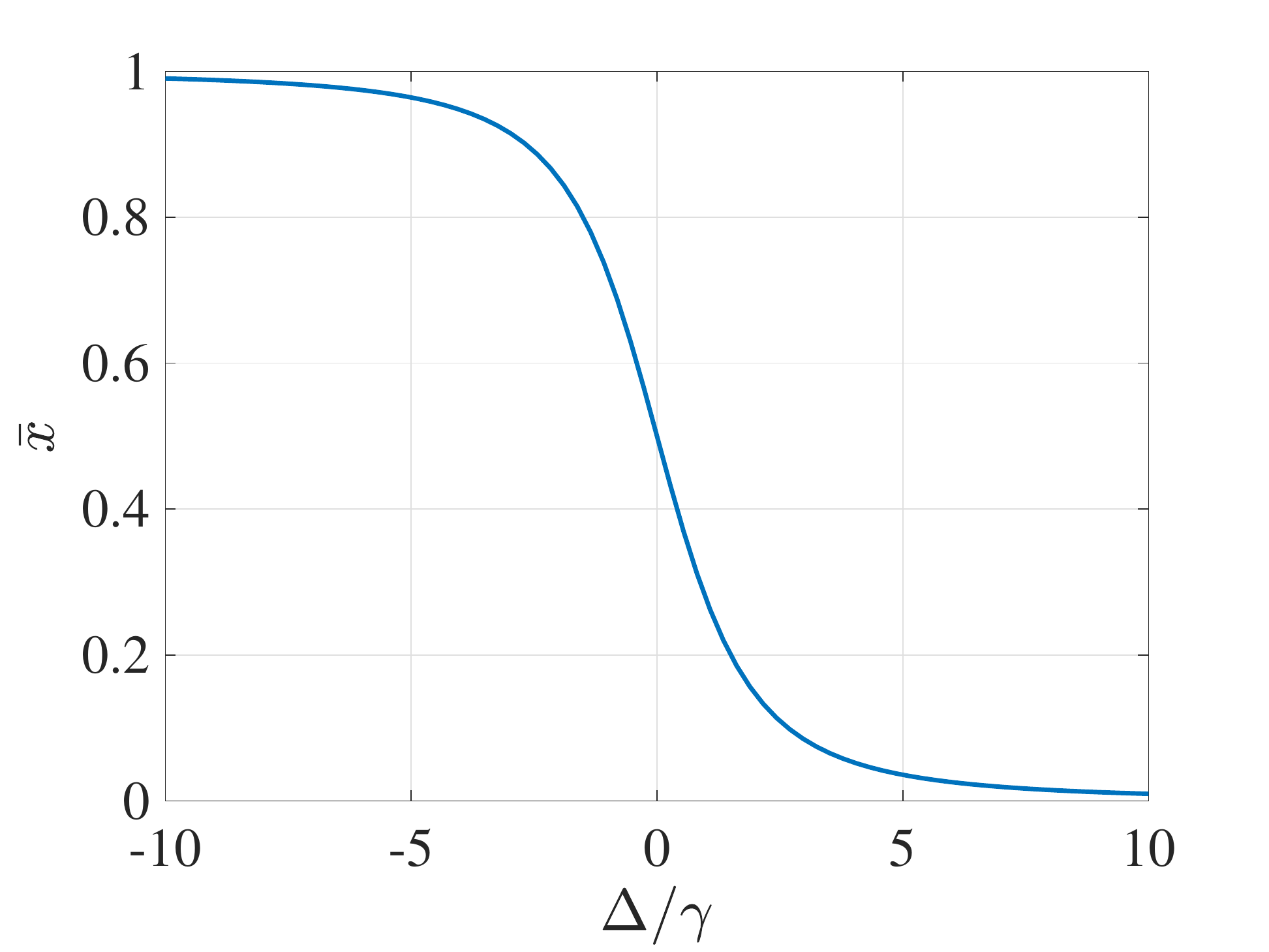} 
   \caption{The mean value of measurements, $\bar{x}$, is tunable by adjusting the detuning $\Delta$ for a pair of coupled quantum dots. Here, it is assumed that the tunneling energy $\gamma$ between dots is fixed.}
   \label{fig:TunableMeanImplementationAgnostic}
\end{figure}

The mean $\bar{x}$ is tunable by adjusting the detuning $\Delta$ (see Fig.\ \ref{fig:TunableMeanImplementationAgnostic}). In the limit of small detuning ($|\Delta| \ll \gamma$), the DQD is unbiased, and $\bar{x} \rightarrow 1/2$. For a large positive $\Delta$ (i.e., $\Delta/\gamma \gg 1$), the $\ket{0}$ component is dominant in $\ket{\phi_1}$ ($\bar{x} \rightarrow 0$, i.e.\ the DQD is strongly biased toward ``0''), and a large negative $\Delta$ (i.e., $\Delta/\gamma \ll -1$) causes the $\ket{1}$ component to dominate ($\bar{x} \rightarrow 1$, the DQD is strongly biased toward ``1'').  In the implementations discussed in this paper, $\gamma$ is a fixed property of the physical system, and $\Delta$ is a tunable bias parameter. Thus, the tunneling energy, $\gamma$, determines the relevant energy scale for $\Delta$. In Fig.\ \ref{fig:TunableMeanImplementationAgnostic}, the mean value $\bar{x}$ is plotted as a function of $\Delta/\gamma$.

\subsection{Generating a String of Random Bits}

\begin{figure*}[htbp] 
   \centering
   \includegraphics[width=2\columnwidth]{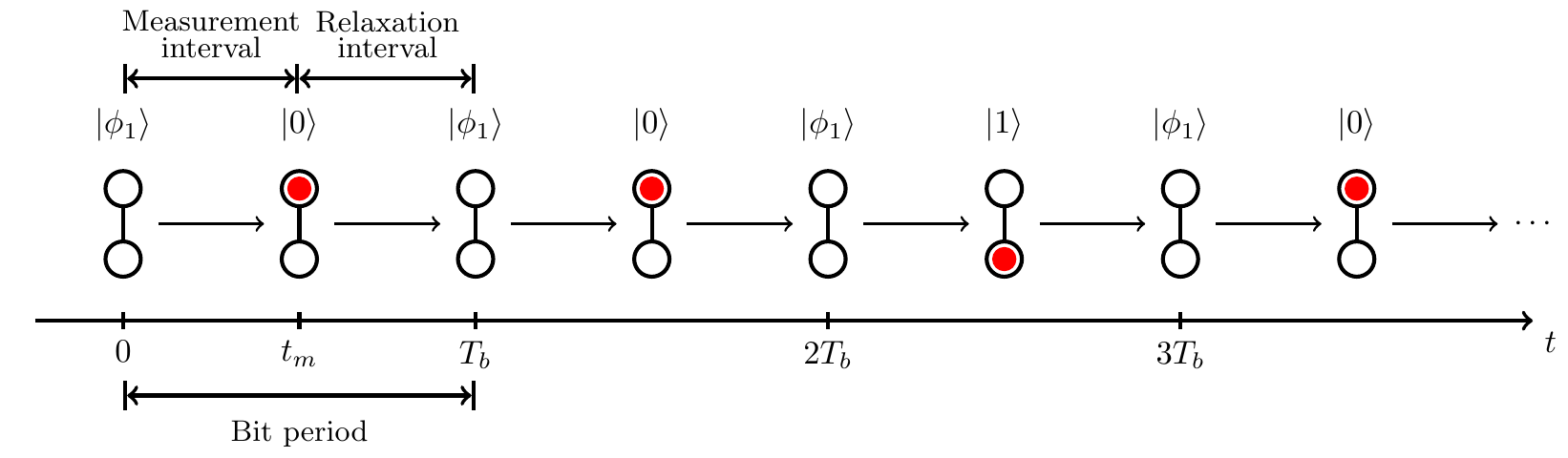} 
   \caption{A series of measurements on a single DQD produces a random bit string ``$0010\cdots$''. A bit is produced in an interval of duration $T_b$, which includes a measurement of interval $t_m$, followed by a relaxation to the ground state over an interval $T_b - t_m$.}
   \label{fig:RandomBitString}
\end{figure*}

Strings of random bits may be generated in parallel using one measurement from each of many DQDs in an array; or in series, by taking many sequential measurements on the same DQD. It will likely be preferable to use the serial method, as the interconnection network within a parallel DQD QRNG system will be vastly more complex than that of a serial QRNG. In the serial implementation, one quantum device is associated with each random bit string. 

\begin{figure}[htbp] 
   \centering
   \includegraphics[width=1\columnwidth]{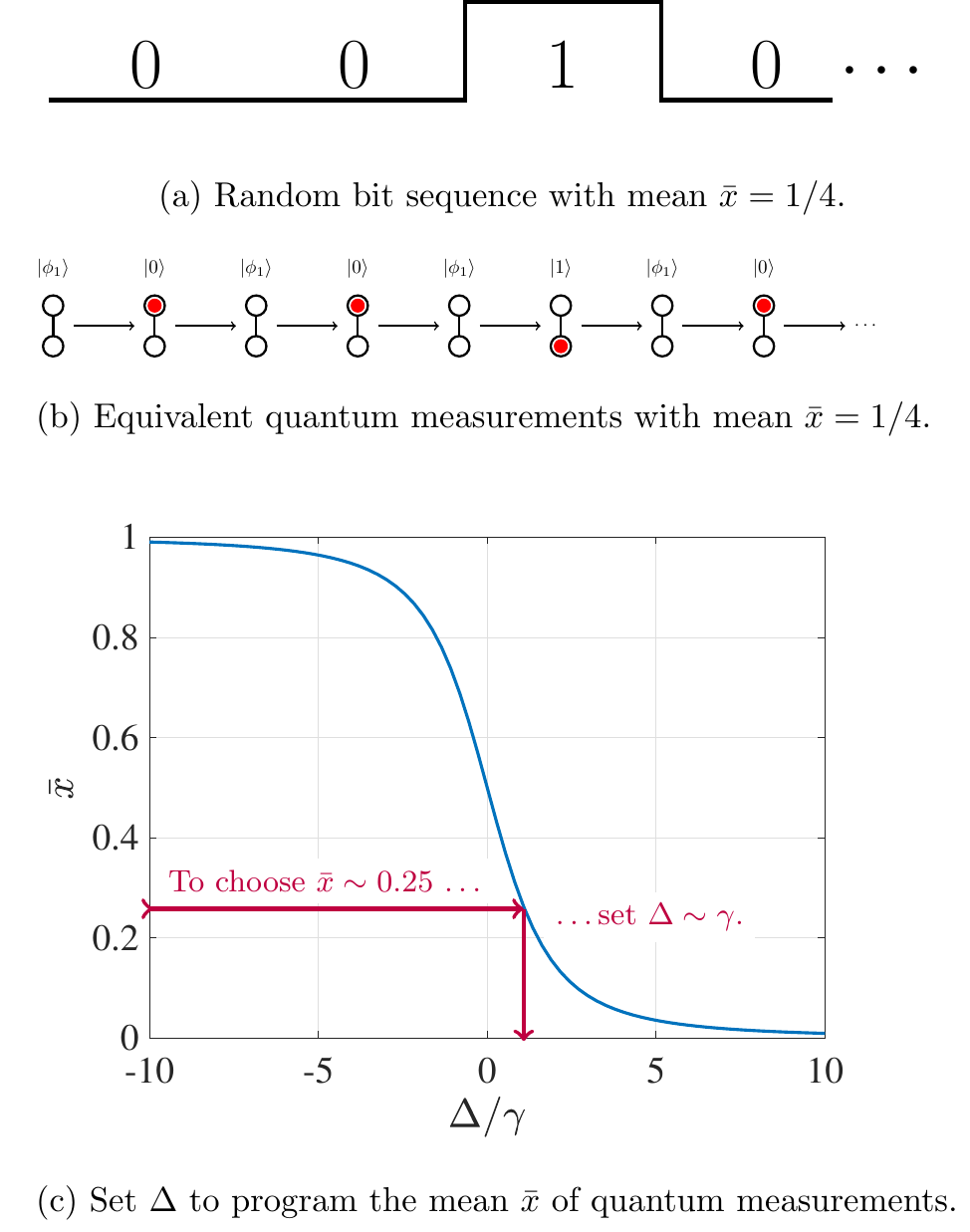} 
   \caption{The mean of random bit measurements is tuned by setting the detuning. (a) A bit string with a desired mean may be implemented by repeated quantum measurements, as shown in (b). The desired mean is programmed by setting the detuning, $\Delta$. This is chosen by selecting $\bar{x}$ on the $y$-axis of Figure \ref{fig:TunableMeanImplementationAgnostic} and then finding the appropriate value of $\Delta$, as shown in panel (c).}
   \label{fig:MeanTuning}
\end{figure}

The serial process for generating a random bit string relies on a measure-relax cycle, as illustrated in Fig.\ \ref{fig:RandomBitString}. A measure-relax cycle requires a time duration $T_b$, which we call the bit time. The first measure-relax cycle begins at $t=0$, at which time the DQD is initially prepared in the ground state $\ket{\phi_1}$. A position measurement is then taken over a duration $t_m$. This is a projective measurement onto either the $\ket{0}$ or $\ket{1}$ state. To complete the measure-relax cycle, the DQD must again be restored to the ground state $\ket{\phi_1}$, and this necessarily requires the dissipation of energy to the environment in a $\ket{0} \rightarrow \ket{\phi_1}$ or $\ket{1} \rightarrow \ket{\phi_1}$ transition. This relaxation may be as simple as waiting for the energy to dissipate. The relaxation back to $\ket{\phi_1}$ completes the measure-relax cycle, and a measurement commences at time $T_b$ to begin the second cycle. $N$ cycles are completed to generate an $N$-bit random sequence. The mean value, $\bar{x}$, is tuned by adjusting $\Delta$, as illustrated in Figure \ref{fig:MeanTuning}.

Figure \ref{fig:MeanTuning} must be considered with the following caveat: for a small set of measurements (small $N$), it is possible for the mean of measurements, $\bar{x}$, to be very different from the desired mean. This is because the probabilities given in Equations (\ref{eqn:px}), (\ref{eqn:p0}), and (\ref{eqn:p1}) are understood to describe ensemble measurements (large $N$). For small $N$, however, the measured $\bar{x}$ only approximates the desired mean. As $N$ grows large, the measured $\bar{x}$ will approach the programmed mean. Here, an accuracy-versus-speed tradeoff arises: for minimum error between the measured $\bar{x}$ and the programmed value, a larger set of measurements (larger $N$) is needed, increasing the length of the bit string along with its acquisition time, $NT_b$.

Because generation of a bit string will be a serial operation, the relaxation time $T_1$ for the quantum system will play an important role in defining the bit rates possible for generating a random bit string as well setting the time scale for single measurement operations. Also, it is assumed that the bit time $T_b$ is much longer than the time scale of the dynamics of electron transfer in the DQD ($T_b \gg \pi \hbar/\gamma$) so that repeated measurements in the measure-relax cycle do not ``freeze'' the DQD in a measured state via the quantum Zeno effect.

\subsubsection{Minimum Sampling Period}

Since serial measurements will be used to generate the stochastic bit sequence, the relaxation time $T_1$ characteristic of the DQD will determine $T_{b,\mbox{\tiny{min}}}$, the minimum $T_b$ for the quantum system, and thus the upper limit of the bit generation rate, $1/T_{b,\mbox{\tiny{min}}}$. Each serial measurement on the DQD in its ground state $\ket{\phi_1}$ will project its state onto one of the localized states of basis $\mc{B}$, neither of which are the ground state in general. Therefore, in order for the DQD to reset from the measured state $\ket{x}$ to the desired ground state $\ket{\phi_1}$ for a subsequent measurement, the system will require some minimum $T_{b,\mbox{\tiny{min}}}$, where $T_{b,\mbox{\tiny{min}}} > T_1$. Of course, $T_1$\textemdash and thus $T_{b,\mbox{\tiny{min}}}$\textemdash is determined by the physical implementation of the DQD system.

\subsubsection{Maximum Sampling Duration}
The relaxation time $T_1$ also determines an upper limit on the duration of a sampling process. During the process of acquiring one random bit, it is necessary to limit measurement time $ t_m $ to $t_m \ll T_1$. This is necessary so that during measurement, the measured state $\ket{x}$ cannot relax back to the ground state and be projected again onto $\mathcal{B}$. If this were to happen, then each measurement itself could be an averaged measurement rather than an individual sample.

\subsection{Power Dissipation}
Power dissipation is a necessary consequence of the ``relax'' portion of the random bit generation process. The measurement projects the system onto $\ket{0}$ or $\ket{1}$, with energies $\braket{0 | \hat{H} | 0}$ and $\braket{1 | \hat{H} | 1}$, respectively. Thus, the energy $ \varepsilon_1 = \braket{1 | \hat{H} | 1} - E_1$ or $\varepsilon_0 = \braket{0 | \hat{H} | 0}  - E_1$ is dissipated in the relaxation between measurements. 
The occupation energies of the states $\ket{0}$ and $\ket{1}$ are $\braket{0|\hat{H}| 0} = - \Delta/2$ and $\braket{1|\hat{H}| 1} = \Delta/2$ so that the energies of relaxation are:
\begin{align*}
\varepsilon_0 & = \frac{1}{2}\left(\sqrt{4\gamma^{2}+\Delta^{2}}-\Delta\right), \; \mbox{and} \\
\varepsilon_1 & = \frac{1}{2}\left(\sqrt{4\gamma^{2}+\Delta^{2}}+\Delta\right) \; .
\end{align*}
Thus, the average energy dissipation $\bar{E} \TextSub{diss}$ over many measure-relax cycles is
\begin{equation} \label{eqn:EnergyDissipation}
\bar{E} \TextSub{diss} = p \Argum{0} \varepsilon_0 + p  \Argum{1} \varepsilon_1  = \frac{2 \gamma^2 }{\sqrt{4 \gamma^2 + \Delta ^2}}.
\end{equation}
Then, over a given time interval of $\Delta t$ in which $N$ measure-relax cycles have been completed, the bit rate is $N / \Delta t$, and the average power dissipation may be calculated as
\begin{equation}
\bar{p} \TextSub{diss} = \frac{N 2 \gamma^2 }{\Delta t \sqrt{4 \gamma^2 + \Delta ^2}}.
\label{eqn:power_dissipation}
\end{equation}
This implies that maximum power dissipation occurs for an unbiased system ($\Delta = 0$). In this case, power dissipation is $N \gamma /\Delta t$. Maximum power dissipation may be decreased by reducing the tunneling energy $\gamma$ or by reducing the bit rate $N/ \Delta t$.

\section{Device Implementations}

Two DQD-based RNGs are discussed here. In the above discussion of device theory, the randomness is provided by Born's rule and the collapse of the wave function. This is directly applicable to the molecular device implementation discussed in subsection \ref{subsect:molecularQD}. In subsection \ref{subsect:lithographicDQD}, an analogous implementation in lithographic quantum dots is discussed. In this case,  statistical mechanics underlies the randomness of the system.

\label{sect:implementations}

\subsection{Molecular Quantum Dot Systems} \label{subsect:molecularQD}
\begin{figure}[htbp] 
   \centering
   \includegraphics[width=1\columnwidth]{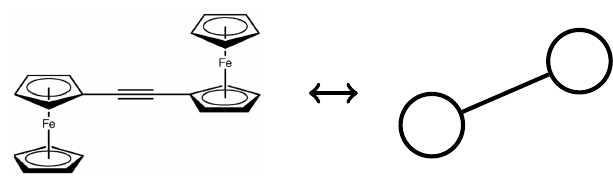} 
   \caption{The two ferrocene groups of a diferocenyl acetylene molecule provide a molecular double-quantum dot (DQD) system.}
   \label{fig:DFA_QCA}
\end{figure}

Molecular DQDs have been conceived for applications such as molecular charge qubits \cite{2013MolecularChargeQubit} and room-temperature, low-power classical computing devices known as quantum-dot cellular automata \cite{LentTougawPorodBernstein:1993, LentScience2000}. Here, a single mixed-valence molecule provides a coupled pair of dots, with redox centers functioning as quantum dots. An example of this is diferrocenyl acetylene (see Fig.\ \ref{fig:DFA_QCA}), an organometallic molecule. A DFA molecule has two ferrocene groups, each providing one quantum dot.

In one system we proposed for biasing a molecular DQD using an applied electric field \cite{QCAeFieldWriteIn-IEEE}, the detuning is
\begin{equation}
\Delta = -q_e \vec{E} \cdot \vec{a},
\end{equation}
where $q_e > 0$ is the fundamental charge, $\vec{E}$ is an applied biasing electric field, and $\vec{a}$ is the vector of length $a$ in a direction pointing from dot 0 to dot 1. 
Here, $a = || \vec{a} ||$ is the distance between the coupled dots. Thus, tuning $\Delta$ is a matter of adjusting the strength of the applied $\vec{E}$. In Ref.\ \cite{QCAeFieldWriteIn-IEEE}, the field is applied using charged electrodes, so the field strength may be changed by varying the voltage $V$ applied between the electrodes.

For a DQD based on an ionic DFA molecule, $a = 0.67~\mbox{nm}$, and it has been calculated that the relaxation time is $T_1 \sim 1~\mbox{ps}$ and the tunneling energy is $\gamma \sim 50~\mbox{meV}$\cite{molecularQCAelectronTransfer}. This results in a maximum measurement frequency of $1~\mbox{THz}$. At this rate, maximum power dissipation will be $8~\mbox{nW}$. 

Bit read-out could be accomplished using single-electron transistor (SET) electrometers, which have demonstrated sensitivity to sub-nanometer displacements of single electrons \cite{SETSensitivitySnider}. While the time scale of molecular device operation is of the ps scale, SET electrometers will likely restrict measurement periods to the time scale of ns. Since this reduces the bit rate $N/\Delta t$, power dissipation is limited as direct a consequence of  Equation (\ref{eqn:power_dissipation}).

\subsection{Lithographic Quantum Dots} \label{subsect:lithographicDQD}

Systems of quantum dots with localized charge states have been fabricated and tested for application in a low-power, classical computing paradigm known as quantum-dot cellular automata (QCA) \cite{LentTougawPorodBernstein:1993,LentTougawArchitecture}. QCA devices have been implemented using both metallic quantum dots \cite{Orlov1997,clockedShiftRegMetalDot} and semiconductor systems \cite{Gardelis2003,Smith2003}. The DQD tunable QRNG may be designed in a similar manner. 

In particular, this discussion focuses on a metal-dot QRNG, which is an analog implementation of the system described in Subsection \ref{subsect:molecularQD}. We make this distinction because here because the electronic system is comprised of many electrons in the solid state device, and there is no coherent single-electron wave function. Just as before, a measurement will result in the electron occupying either state ``0'' or ``1''; however, the randomness of the outcome is generated not by Born's rule and the collapse of a wave function, but rather by the Boltzmann statistics of random electron transfer events between the ``0'' and ``1'' states within the DQD in thermal equilibrium with its environment. Here, the ``relax'' part of the measure-relax cycle in the generation of an $N$-bit string is more appropriately called a ``randomization'': as time elapses within the randomization phase, thermal fluctuations may drive an accumulating series of random, interdot electron transfer events. This again give rise to probabilities of measurement $p \Argum{x}$ that are functions of the detuning $\Delta$ and rates of electron transfer for $\ket{0} \rightarrow \ket{1}$ and $\ket{1} \rightarrow \ket{0}$ transitions.

In a lithographic implementation, the detuning $\Delta$ may be achieved by directly applying a voltage between dots 1 and 0: $\Delta = q V$, where $q$ is the mobile charge, and $V$ is the voltage between the dots. Thus, an adjustment to $\Delta$ is made by varying $V$. State read-out may be achieved by measuring the charge state using SET electrometers.

In this implementation, cryogenic cooling is required. Such devices could be deployed within quantum computers, where operation typically is at cryogenic temperatures. Here, a metal-dot QRNG could provide native, hardware-based random numbers.

We perform some simple calculations for a system of coupled quantum dots as presented by Amlani, \textit{et al} \cite{Orlov1997}. For this type of cell, the relaxation time was estimated at $T_1 \sim 7 ~\mbox{ns}$ \cite{QCA_qbits}, which corresponds to a maximum bit rate of $150\times 10^{6}$ bits/s (or a sampling rate 150 MHz). If we take $\gamma = 0.5~\mbox{meV}$, then at this rate, maximum average power dissipation is 12 fW.

Table \ref{table:CompareFOM} lists estimated figures of merit for both the metal-dot and molecular implementations of the proposed tunable QRNG. Maximum power dissipation is calculated by using Eqn.\ \ref{eqn:power_dissipation} with zero detuning. For reference, these are compared to estimates for a stochastic number generator (SNG) based on a 32-bit linear feedback shift register (LFSR) synthesized in a 65-nm CMOS process operating at 100 MHz \cite{TRNG_memristor}, with the caveat that the SNG figures of merit are not necessarily for maximum operating speeds. It is noteworthy that in general, smaller DQDs allow higher maximum bit rates.

\begin{table}[htbp]
\caption{Figures of merit for a tunable QRNG design in metallic quantum dots and a diferrocenyl acetylene (DFA) molecular DQD system.}
\begin{center}
\begin{tabular}{|p{1.5cm} | p{1.5cm} |p{1.5cm}|| p{1.75cm} |}
\hline
\textbf{Figure of Merit} & \textbf{Metallic DQD} & \textbf{Molecular DQD} & \textbf{CMOS SNG} \cite{TRNG_memristor} \\
\hline 
$T_1$ & $6.67~\mbox{ns}$ & $\sim 1~\mbox{ps}$ & -  \\
\hline
Maximum bit rate (bps) & $150 \times 10^{6}$ & $1 \times 10^{12}$  & $100\times10^{6}$ \\
\hline
Maximum average power dissipation (W) & $12 \times 10^{-15}$ & $8 \times 10^{-9}$ & $80.2 \times 10^{-6}$ \\
\hline
\end{tabular}
\end{center}
\label{table:CompareFOM}
\end{table}%

\section{Conclusion}

We have proposed a straight-forward, minimalistic hardware approach to a tunable quantum random number generator. A random bit is generated by taking a position measurement of the ground state of a mobile charge on a coupled pair of quantum dots. A random bit string is simply a sequence of such measurements, the mean value of which may be tuned by varying the bias of the DQD system. We discussed both a molecular impementation of the DQD, as well as a metallic implementation. Such devices could be integrated in quantum computing or quantum communication systems. Also, the proposed QRNG could provide a low-power, hardware-based approach to a tunable stochastic number generator for stochastic computing.

\section*{Acknowledgment}

The authors thank Craig S.\ Lent of the University of Notre Dame for discussion on the topics of quantum systems and measurement as applied to this concept. This work was supported by Baylor University under a new-faculty start-up grant.

\ifCLASSOPTIONcaptionsoff
  \newpage
\fi



\bibliographystyle{IEEEtran}
\bibliography{DQD_RNG_refs}
\end{document}